\documentclass[reprint,fleqn,superscriptaddress,twocolumn,showpacs,
amsmath,amssymb,prb,aps]{revtex4-2}
\usepackage[english]{babel}
\usepackage[all]{xy}
\usepackage{graphicx,psfrag,times,epsfig,color}
\usepackage{verbatim,natbib}
\usepackage{booktabs}
\usepackage{color}
\usepackage{makeidx}
\usepackage{amsmath}
\usepackage{bm}
\usepackage{amsfonts}
\usepackage{amssymb}
\usepackage{hyperref}
\usepackage{bm}
\usepackage{ragged2e}

\begin{document}

\let\namerefOld\nameref
\renewcommand{\nameref}[1]{\textit{\namerefOld{#1}}}

\title{Common superconducting transition in under and overdoped cuprate superconductors}
\author{H\'ercules H. Santana and E. V. L. de Mello}
\affiliation{Instituto de F\'{\i}sica, Universidade Federal Fluminense, 24210-346 Niter\'oi, RJ, Brazil}


\begin{abstract}

Underdoped cuprate superconductors are believed to be strongly correlated with electronic systems with
small phase stiffness leading to a large
phase fluctuation region is known as the pseudogap state. With increasing doping it is generally agree
that they become Fermi liquid, rendering the end of the superconductivity
due to the sufficiently large electronic screening. However, this scenario does not
stand against a recent experiment\cite{OverJJ2022} that combined 
magnetic susceptibility and Scanning Tunnelling Microscopy (STM)
which measured superconducting gaps and amplitudes amid charge inhomogeneity
far beyond the critical doping $p_{\rm c} \approx 0.27$.
We reproduced these results by calculating the localized superconducting
amplitudes that emerge out
of charge inhomogeneities, which forms a mesoscopic granular superconductor with an
array of Josephson junctions, whose average couplings determine the critical temperature $T_{\rm c}$.
The calculations agree with the experiments and both
yield that underdoped and overdoped compounds have superconducting
long-range order by the same mechanism.

\end{abstract}
\maketitle

\section{Introduction}

Cuprates superconductors are formed by doping holes in an
antiferromagnetic insulator, and earlier experiments\cite{Huefner2008,Keimer2015} indicated that 
underdoped and overdoped materials were quite different.
The underdoped regime exhibits the pseudogap phenomenon below the characteristic 
temperature $T^*$,
as well as tendencies towards several incipient orders, including charge 
density waves (CDW) or incommensurate charge order (CO).
The pseudogap and superconducting gap have both the same
$d$-wave symmetry\cite{Lee2007,Kanigel2008} suggesting that the 
latter is from preformed pairs without long-range phase coherence, 
but their precise relationship is still a matter of intense debate\cite{Lee2007,Keimer2015}.
The temperature $T^*$ and the CDW/CO signals, in general, 
decrease with doping\cite{Huefner2008} 
which led to models that overdoped compounds were homogeneous 
and possess Fermi liquid properties. This scenario 
was supported by some experiments like, for instance, 
quantum oscillations in overdoped\cite{Vignolle2008} Tl$_2$Ba$_2$CuO$_{6+\delta}$. 

Along this line, Emery and Kivelson\cite{Emery1995} proposed that $T_{\rm c}$ of underdoped
cuprates are constrained by the small superconducting phase stiffness, 
rather than by the onset of pairing correlations,
which was consistent with the observation of preformed pairs\cite{Kanigel2008}. 
On the other hand,
they assumed that the decrease of $T_{\rm c}$ with overdoping
would be controlled by the mean-field behavior predicted by the BCS-Eliashberg theory\cite{Emery1995}.
While their theory was generally accepted, some more recent experiments
contradicted their model. For instance,
Nernst and torque magnetization experiments have provided evidence that
the disappearance of the zero resistivity state at $T_{\rm c}$ is caused by a lack of
long-range phase order\cite{Nernst2010,Rourke2011,NernstPRB2018} rather than by the vanishing of the
superconducting gap. A result that is also in agreement with scanning tunnelling
microscopy (STM) on Bi-based cuprates\cite{Gomes2007} and with muon
spin rotation experiments ($\mu$SR)\cite{Muon2013}, which did not observe any 
particuar change of behavior at and across $T_{\rm c}$.

To elucidate how superconductivity disappears at large dopings, a set of
experiments were performed\cite{OverJJ2022}: starting with
susceptibility measurements on La$_{2-x}$Sr$_x$CuO$_4$ (LSCO)
with $x = p = 0.25$, they found an onset of diamagnetism
at $T_{\rm c1} = T_{\rm c}^{on} = 38.5$ K
and a sharp transition to a Meissner state at $T_{\rm c2} = T_{\rm c} = 18$ K.
These results revealed a large interval of superconducting
fluctuation. Accordingly, they studied the average compound $\left < p \right >= 0.29 $ that
is beyond the superconducting dome, and surprisingly, both samples show the same
onset of diamagnetism near 38 K; however, for $p = 0.29$, the development of
a bulk-like response only occurs near 5 K. To refine their findings,
they also performed spectroscopic imaging
Scanning Tunnelling Microscopy (SI-STM) on a
metallic LSCO film with $p \approx 0.35$, which provided indications that, even beyond the
superconductor-to-metal transition, there are small regions possessing pairing gaps, that survive well above
$T_{\rm c}$ in an electronic inhomogeneous media. It implies that local pairing
amplitudes are more easily formed in the presence of charge disorder like
puddles but, as they pointed out, the amplitudes are too dilute to support
superconducting order\cite{OverJJ2022}. They also pointed out that these
results are consistent with the theoretical predictions\cite{Spivak2008}
in which superconductivity initially develops in
isolated grains characterized by a reduced hole concentration,
and bulk superconductivity is induced by the proximity effect.

To test these ideas, two experiments focused mainly on how the superconducting phase
vanishes obtained different results: angle-resolved photoemission
spectroscopy (ARPES) studies of Bi$_2$Sr$_2$CaCu$_2$O$_{8+\delta}$ show that the
pairing amplitude weakens with doping and completely disappears precisely where
superconductivity ends at $p_{\rm c}$\cite{Valla2020}. They concluded that the
superconducting state is determined primarily by the coupling strength or
pairing amplitude.
The other, high-resolution real-space Scanning Tunnelling
Spectroscopy (STS)\cite{Over2023}
on (Pb,Bi)$ $Sr$_2$CuO$_{6+\delta}$ spanning from the underdoped to large hole doping revealed
a saturation of the spatially averaged pairing amplitude
of $\Delta_{\rm sc} \sim 6$ meV near the superconducting end at $p_{\rm c} = 0.27$. Their
analyses led to a different conclusion, namely, that the vanishing
of the superconducting phase occurs by the emergence of nanoscale
superconducting puddles filled in with free carriers.

Here we point out that these apparent contradicted results
can be reconciled assuming
a situation akin to granular superconductivity provided by mesoscopic
CDW/CO or puddle domains, which we believe to be the most fundamental
property of cuprate superconductors.
The calculations are based on CDW/CO
simulations which are done by the Cahn-Hilliard (CH) nonlinear differential equation and
a Ginzburg-Landau (GL) free energy. This approach describes the phase-ordering dynamics
of an intrinsic electronic phase
separation (EPS) transition, assumed to be common to all
cuprates. The phase separation free energy potential $V_{\rm GL}$ develops energy modulations
which are shown in Fig. \ref{fig1}, starting at the pseudogap temperature $T^*(p)$,
taken as the onset of phase separation,
and at the parental $p = 0$ compound\cite{Anjos2025}. At low temperatures,
it segregates the doped holes in grains or charge domains of rich and poor
local hole densities. Such charge inhomogeneities
may induce low-temperature hole-hole interaction proportional
to the amplitude of $ V_{\rm GL}$\cite{TimeEvol2012,Mello2017,Mello2020a}
that are depicted in Fig. \ref{fig1}. These CDW/CO simulations are followed
by Bogoliubov-deGennes (BdG) self-consistent superconducting amplitude calculations
on such charge domain array, which is general to any finite doped
cuprate\cite{Mello2020a,Mello2023}. As observed before,
the superconducting properties depend on the average
amplitude\cite{Mello2017} $\left < \Delta_d (p, T )\right >$ over
the CuO plane. The average Josephson coupling
$\left < E_{\rm J} (p, T )\right >$ between the local SC order parameters
may lead to long-range order (LRO) or phase-ordering at low
temperatures\cite{deMelloKasal2012,DeMello2012,Mello2017,Mello2020a}.

We emphasize that our model simulates
the charge inhomogeneity and provides a way to calculate the local pairing
amplitude that is proportional\cite{Mello2020a} to $T^*$. It is well known that $T^*$
is very large at low doping\cite{Huefner2008} but may be nonzero for LSCO up to $p \sim 0.35$,
according to the limit of the temperature-independent Hall coefficient\cite{Hall.1994} that
is one of the criteria to $T^*(p)$, which
explains the puddling structure in regions beyond $p_{\rm c}$\cite{OverJJ2022}.
Overall, the phase separation model
provides an interpretation to the conflicting SI-STM\cite{OverJJ2022}, ARPES\cite{Valla2020},
and STS\cite{Over2023} experiments and reproduces
$T_{\rm c}(p)$ and $\left < \Delta_d (p, T )\right >$ that defines the
susceptibilities temperature interval of Ref. \citenum{OverJJ2022} and provides a
unified approach to $T_{\rm c}(p)$.

\section{The Model }

Recently, we studied the 
Hall coefficient $R_{\rm H}(p, T)$ temperature dependence\cite{Anjos2025} assuming
the presence of charge inhomogeneities in the entire doping region
including the antiferromagnetic parent compound, i.e., the undoped $p = 0$ or
$n = 1$ hole per Cu atom. This suggested that antiferromagnetic and CDW/CO
are intertwined, which is proposed to be in the origin of stripes\cite{Keimer2015}.
With doping, the extra holes are segregated mainly in
alternating hole-rich and hole-poor domains, from low\cite{Insul.Stripes2019,Kang2023B} ($p \le 0.05$)
to the far overdoped\cite{Fei2019,Tranquada2021} compounds.

To provide a mathematical description of how the charge inhomogeneities
are formed, we use the time-dependent
nonlinear Cahn–Hilliard (CH) differential equation, well-known
in the theory of metallic alloys\cite{Cahn1958}. The method
generates the charge modulation`s free energy that yields
the above-mentioned hole-rich and hole-poor domains\cite{Mello2020a,Mello2022}.
In this approach, the pseudogap temperature $T^*(p)$ acts as the phase separation
limit between uniform charge density and the CDW/CO instabilities.

In previous works\cite{DeMello2012,Mello2017,Mello2020a},
we described the CH method in detail
and its connection with the cuprates' superconducting properties.
Briefly, just for completeness, the starting
point is the time-dependent phase separation order parameter associated
with the local electronic density, $u({\bf r}) = (p({\bf r}) - p)/p$, where $p({\bf r})$
is the local charge or hole density at a position ${\bf r}$ in the CuO plane and
$p$ is the average doping $\left < p \right >$.
The CH equation is based on the minimization of the Ginzburg–Landau
(GL) free energy expansion in terms of
the conserved order parameter\cite{TimeEvol2012} $u(r, t)$;

\begin{equation}
f(u)= {{\frac{1}{2}\varepsilon |\nabla u|^2 +V_{\rm GL}(u,T)}},
\label{FE}
\end{equation}
where  ${V_{\rm GL}}(u,T)= -A(T) u^2/2 + B^2u^4/4+...$ and $A(T) = \alpha (T_{\rm PS}-T)$, is a
double-well potential that characterizes the electronic phase separation into
hole-rich and hole-poor phases
below $T_{\rm PS}$, assumed to be $T^*(p)$. We use
$\alpha$ and $B =1$, and $\varepsilon$ controls the spatial separation of the
charge-segregated patches that are connected with the experimental\cite{Comin2016}
wavelength $\lambda_{\rm CO}(p)$
and drop the temperature dependence\cite{TimeEvol2012,Mello2017,Mello2020a}.

\begin{figure}[!ht] 
\centerline{\includegraphics[height=5.50cm]{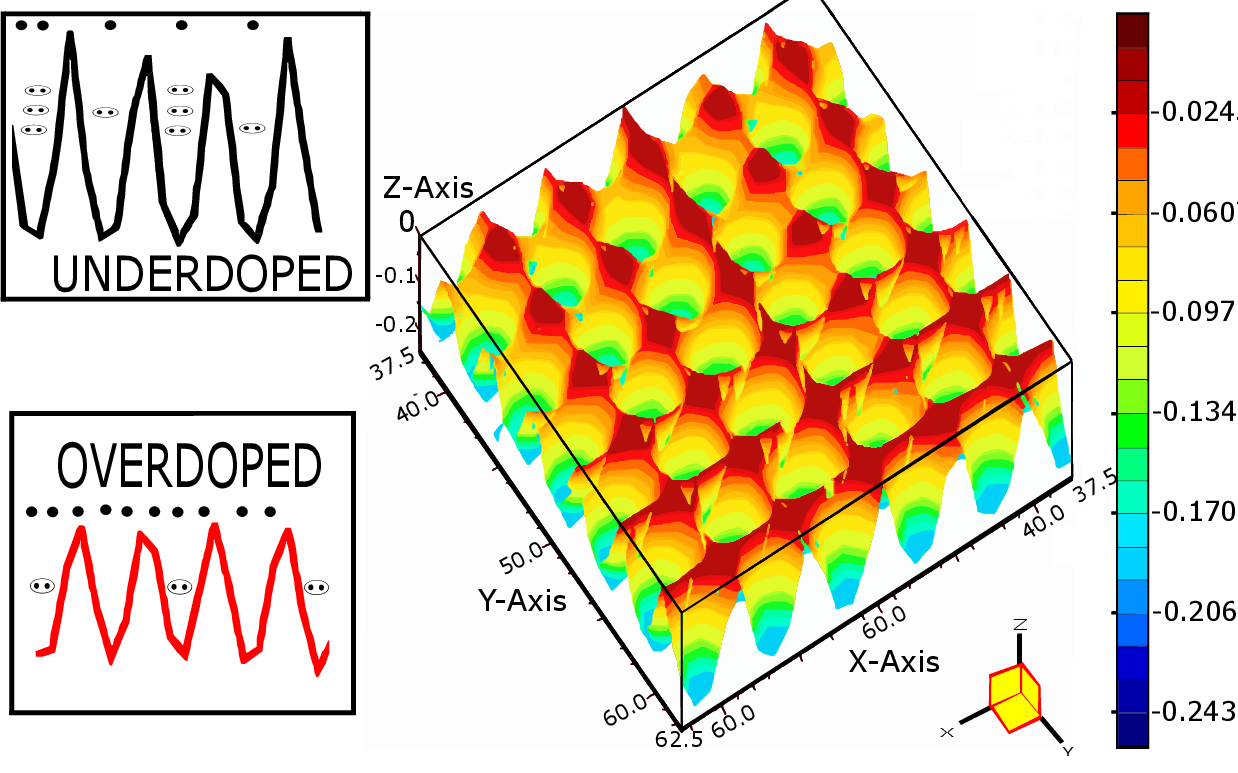}}
\caption{
The CH simulations of the phase separation GL free energy
potential $V_{\rm GL}(u({\bf r}))$ that induces and shapes the
CDW/CO in a central portion of a N x N unit cells square representing the holes at a position $r_i$.
$V_{\rm GL}(r_i)$ along the $x$ direction
in the CuO plane in an overdoped in the top inset and an underdoped sample in the down inset.
These insets show that $V_{\rm GL}$
modulations have larger amplitudes in the
underdoped regime because the EPS scales\cite{Anjos2025} with $T^*$.
}
\label{fig1}
\end{figure}

In Fig. \ref{fig1} we show schematically how the CH phase separation is developed.
We simulate the CDW/CO in a N x N square representing the holes at a position $r_i$
and the GL free energy
potential $V_{\rm GL}(u({\bf r}))$ that shapes the CDW/CO,
depicted in Fig. \ref{fig1}. Going along the $x$-direction,
we can see how the high and low $V_{\rm GL}$ values host alternating rich and poor domains
at low temperatures as shown in Fig. \ref{fig1} in the top and down insets.
We have already argued\cite{TimeEvol2012,deMelloKasal2012}
that this inhomogeneous charge unbalance, favors pairing attraction
inside the charge domains, which is
proportional to the mean $V_{\rm GL}(u({\bf r}))$ amplitude of oscillations $\left < V_{\rm GL} \right >$ shown in
Fig. \ref{fig1} top and bottom insets. $\left < V_{\rm GL} \right >$ is proportional to
the $T^*(p)$ and weakens with doping\cite{TimeEvol2012,Mello2017,Mello2020a} but they may be
non-vanishing even beyond the superconducting limit. Recalling that the Hall
amplitude becomes temperature-independent
near to $p \sim 0.35$ and this is one of the criteria to define
the pseudogap\cite{Hall.1994} $T^*(p)$.

Using the GL phase separation theory, we write the attractive Hubbard
potential as\cite{Mello2020a} $V (p, T) = V_0 \times [1- T / T^*]^2$,
where $V_0 \propto \left< V_{\rm GL} \right >$ is a parameter used to
derive the low-temperature average superconducting amplitude
$\left<\Delta_{\rm sc}(p, 0)\right>$. This is done by the BdG
self-consistent $d$-wave superconducting calculations with
parameters derived for overdoped LSCO from Ref. \onlinecite{Zhong2022}
that are listed in Table 1 of supplementary information (SI).
According to the experiments\cite{Comin2016} the CDW/CO patterns do not
change appreciably with $T$ near the superconducting phase, therefore,
our BdG method consists in keeping a given CO low-temperature structure constant by varying
the local chemical potential at each self-consistent
interaction while $p ({\bf r}_i)$ and $\Delta_{\rm sc}({\bf r}_i)$ are converging\cite{Mello2020a}.
Typical $p ({\bf r}_i)$ and $\Delta_{\rm sc}({\bf r}_i)$ are
shown in Fig. \ref{fig025}.

Fig. \ref{fig025} also shows that $\Delta_d({\bf r}_i,p, T)$ have local spatial 
variations and because the short superconducting coherence length, each 
region with a finite amplitude may be regarded as a grain. Therefore, as in granular superconductors, it gives rise to Josephson coupling between the alternating
charge regions with energy $E_{\rm J}(r_{\rm i j})$, which is the lattice version
of the local superfluid density\cite{Spivak1991} $\rho_{ sc} (r{\rm i j})$.
Which imply that global properties\cite{Mello2021,Mello2022} like
the condensation energies, critical temperatures, and interlayer Josephson
coupling are function of the average SC amplitudes
$\left <\Delta_d(p,T)\right > = \sum_{i}^{N} \Delta_d (r_i, p, T)/N $, where $i$ runs
over $N$ unit cells of the CuO plane. 
Based on the work of Bruder {\it et al}\cite{Bruder95}, even for $d$-wave amplitudes, 
it is sufficient to use the Ambegaokar-Baratoff analytical $s$-wave expression\cite{AB1963}
averaged over the plane:
\begin{equation}
 {\left < E_{\rm J}(p,T) \right >} = \frac{\pi \hbar {\left <\Delta_d(p,T)\right >}}
 {4 e^2 R_{\rm n}(p)} 
 {\rm tanh} \bigl [\frac{\left <\Delta_d(p,T)\right >}{2k_{\rm B}T} \bigr ] .
\label{EJ} 
\end{equation}
Where $R_{\rm n}(p, \sim T_{\rm c})$ 
is proportional to the corresponding normal state resistance just above $T_{\rm c}$. 
Notice that ${\left < E_{\rm J}(p,T) \right >}$ is a global property and may be 
nonzero even when 
$ \Delta_d (r_i, p, T)$ vanish in some regions, provided that the averages 
$\left <\Delta_d(p,T)\right >$ is
finite. By the same token, ${\left < E_{\rm J}(p,T) \right >}$ can be vanishing small with a dilute
superconducting amplitude occupying a small volume fraction, which we believe 
is the situation of
the far overdoped compounds of Refs. \onlinecite{OverJJ2022} and  \onlinecite{Over2023}.

\section{Results}

\begin{figure}[!ht] 
\centerline{\includegraphics[height=3.70cm]{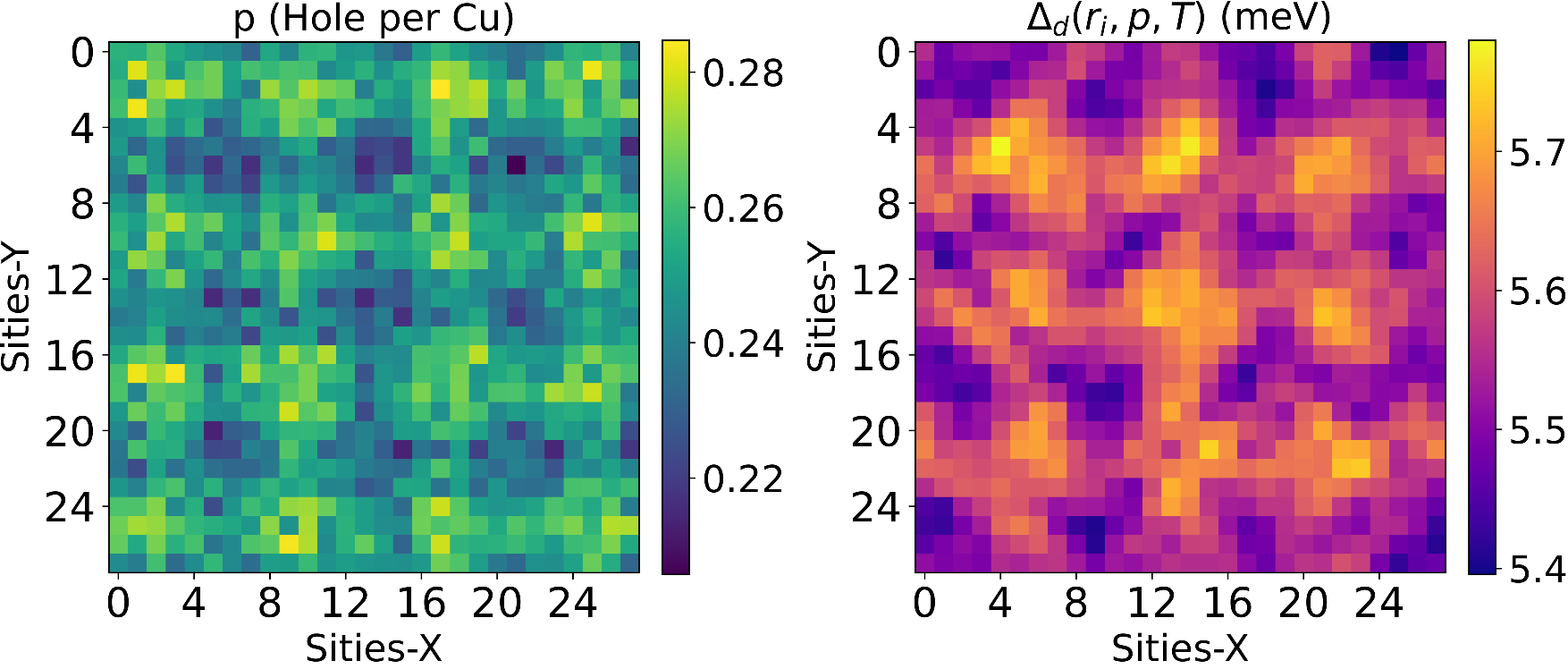}}
\caption{In the left we show the local variations $p(r_i)$ of a compound with average
doping $p = 0.25$. In the right, the two-dimensional BdG $\Delta_d(p=0.25,0)$ calculations on the same
location. It demonstrates that, everything being the same, local larger densities have smaller local superconducting
amplitudes and this points towards why the superconducting gaps vanish in strong overdoped regions.
}
\label{fig025}
\end{figure}

Now, we apply these calculations to compounds near the superconducting-to-metal
experiments mentioned in the introduction.
We start with the measurements of Ref. \onlinecite{OverJJ2022} on
a LSCO $\left < p \right > = p = 0.25$ compound and simulate stripe-like charge density oscillations
that are shown in Fig. \ref{fig025}. On this system we calculate the BdG superconducting amplitude map
$ \Delta_d (r, T=0)$ that are shown in the down inset. The calculations follow up
with the average $\left < V_{\rm GL}(p)\right >$ giving an attractive potential $V_0$ in such way that
$\left <\Delta_d(p = 0.25, T=0)\right >$ oscillates around $\sim 5.6$ meV, that is shown in the
right of Fig. \ref{fig025}, which is in
agreement with the low temperature measurements of Kato {\it et al}\cite{Kato2008}.
After this,
we perform finite temperature calculations $\left <\Delta_d(p,T)\right >$ displayed in
Fig. \ref{fig025} that, with Eq. \ref{EJ}, is used to
derive the sample critical temperature. We recall that
$T_{\rm c}(p)$ is the long-range phase order temperature, and it is obtained
by the competition between the
average Josephson coupling between the local phases $\theta_i$ in the charge domains
and the disorder given by the thermal energy $k_{\rm B}T$.
By plotting both together in Fig. \ref{figAllEJ}, we derive $T_{\rm c}(0.25) \approx 19$ K  when they are equal,
that is, ${\left < E_{\rm J}(p,T_{\rm c}) \right >} = k_{\rm B}T_{\rm c}$ by comparing both
sides of Eq. \ref{EJ}, which is plotted in Fig. \ref{figAllEJ}

\begin{figure}[!ht] 
\centerline{\includegraphics[height=5.0cm]{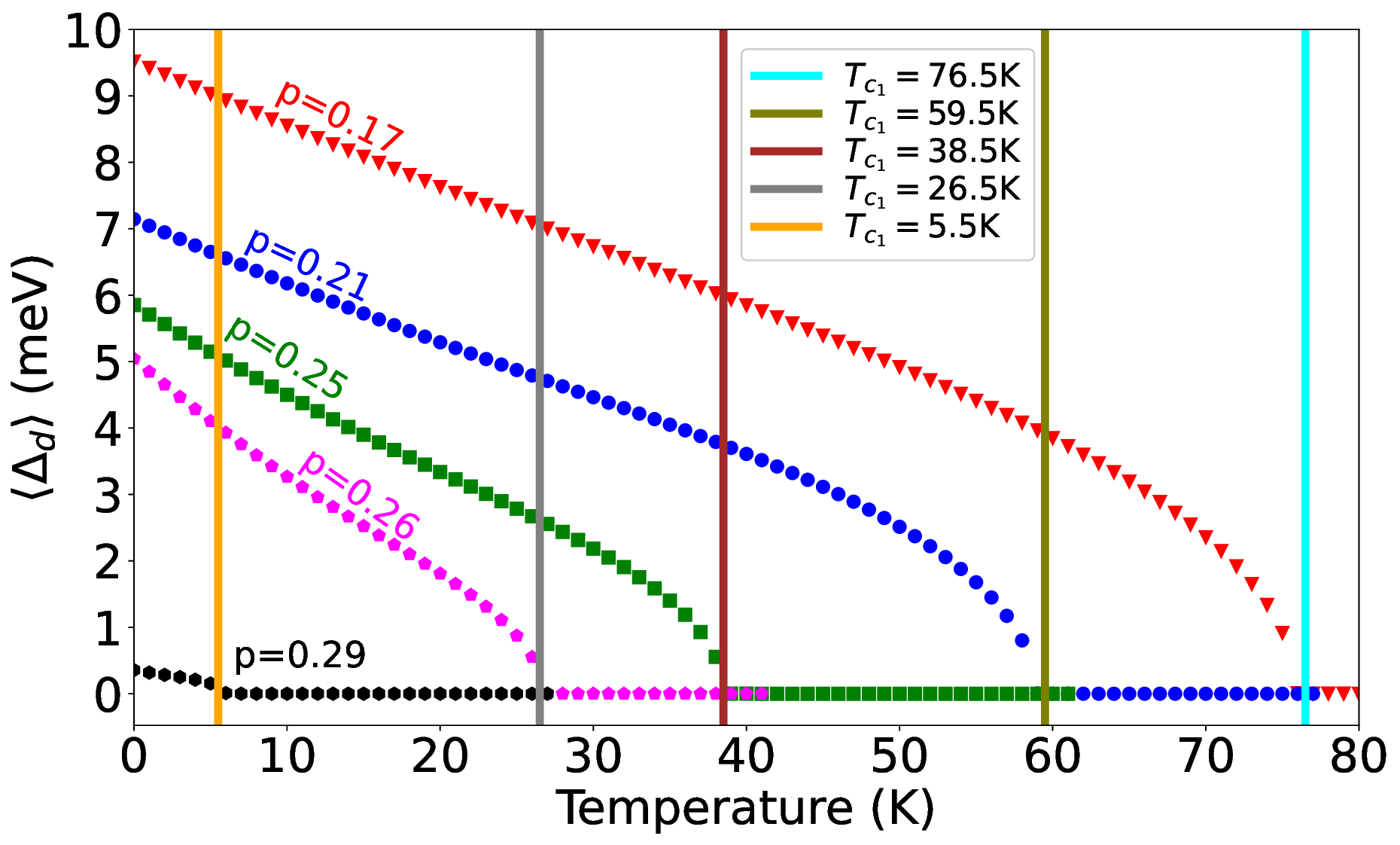}}
\caption{ The average $d$-wave superconducting amplitude $\left <\Delta_{\rm sc}(p, T)\right>$
as function of temperature
for some overdoped compounds. The zero temperature results match the
experimental values\cite{Kato2008} and we derive the onset of superconducting fluctuations $ T_{\rm c}^{on}(p)$
when $\left <\Delta_{\rm sc}(p, T)\right> \rightarrow 0$ that coincides with the
onset of susceptibility signal of Ref. \onlinecite{OverJJ2022} and the Nernst
effect\cite{Nernst2010,Rourke2011,NernstPRB2018}.}
\label{figAllD}
\end{figure}

\begin{figure}[!ht] 
\centerline{\includegraphics[height=5.0cm]{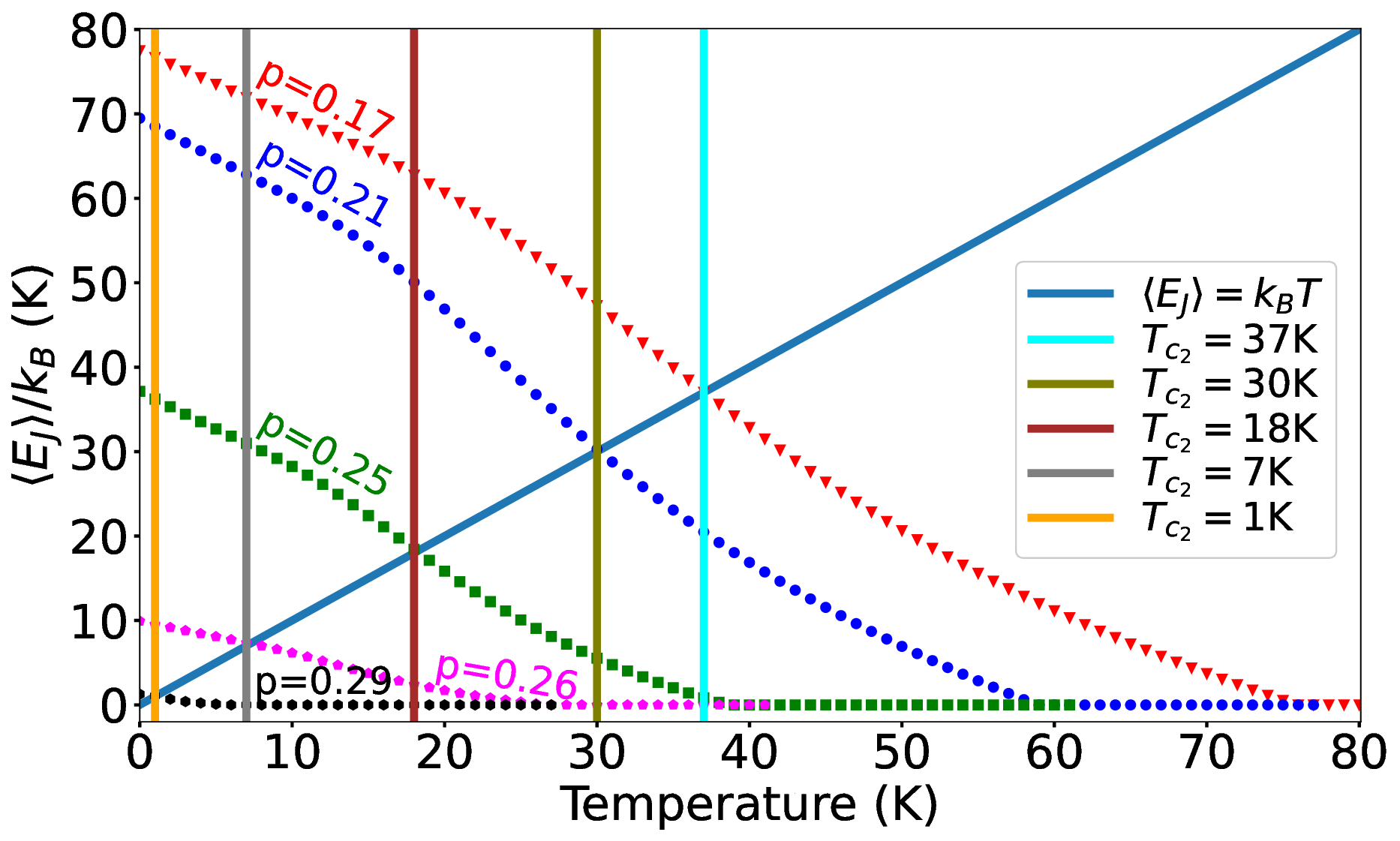}}
\caption{ The average Josephson coupling as function of temperature given in Eq. \ref{EJ}
and the thermal fluctuation energy $k_{\rm B}T$. The
long range phase order transition at $ T_{\rm c}(p)$ occurs
when both energies are equal. That is, when the straight line $k_{\rm B}T$ crosses
the $\left < E_{\rm J}(p,T_{\rm c}) \right > $ curve.
}
\label{figAllEJ}
\end{figure}

In Fig. \ref{figAllD} we show that the average $\left <\Delta_d(p,T)\right >$  used above to
calculate $T_{\rm c}$ vanishes
at $T \approx 40$ K, much above $T_{\rm c}$ but in close
agreement with the measured diamagnetic or susceptibility onset\cite{OverJJ2022}. Another interesting
result from these BdG calculations is that, everything being the same, larger local densities yield lower
local amplitudes that can easily be seen in Fig. \ref{fig025}, in agreement
with theoretical predictions for cuprates\cite{Spivak2008}.

We repeat this procedure to other overdoped compounds\cite{OverJJ2018,OverJJ2022}
$\left < p \right > = p = 0.17, 0.21, 0.26$, and 0.29, starting always with
the zero temperature $\left <\Delta_{\rm sc}(p, 0)\right >$
looking for agreement with the results of Ref. \onlinecite{Kato2008}.
After this, the temperature dependent $\left <\Delta_{\rm sc}(p, T)\right >$
are evaluated using $V (p, T)$ mentioned
above with $T^*(p)$ from the literature, and the results are
plotted in Fig. \ref{figAllD}. Where $\left <\Delta_{\rm sc}(p, T)\right >$ vanishes
determines $ T_{\rm c}^{on}(p)$,
the onset phase-fluctuation region above $ T_{\rm c}$
that was studied by several methods like, STM\cite{Gomes2007} and $\mu$SR\cite{Muon2013}
on Bi-based cuprates, susceptibility\cite{OverJJ2018,OverJJ2022} and Nernst
effect\cite{Nernst2010,Rourke2011,NernstPRB2018} on LSCO.
The parameters used in the Hubbard model of the BdG calculation are in the
Supplemental Material at  [URL will be inserted by publisher] which uses
Refs. \citenum{Hoek2016,Kato2005,Kato2007,Kim2004,Fujimori1998}.

These calculations yield good quantitative results of the measured\cite{OverJJ2018,OverJJ2022,Rourke2011}
phase-fluctuation region between $T_{\rm c}^{on}(p)$ and $T_{\rm c}(p)$
on $p = 0.17, 0.21, 0.25$, and 0.29 compounds, suggesting that the superconducting amplitude is
indeed nonzero above $T_{\rm c}$ and it may also be the case at the end and beyond of superconducting
phase at $p_{\rm c}$. The difference between the zero values
obtained by ARPES\cite{Valla2020} is probably because the real space
STM and STS of Refs. \onlinecite{OverJJ2022} and \onlinecite{Over2023} probe essentially the CuO
unit cell and obtain local gaps while the $k$-space technique
probes the average over small spatial regions.

\section{Conclusion}

We have mentioned that there is not, so far, a consensus of how superconductivity 
disappears in cuprates at the large hole doping limit $p_c$. 
We studied this problem with a
model based on a mesoscopic granular superconductor applied previously to any finite doping\cite{Mello2020a} to
recent experiments with overdoped LSCO compounds near the superconductor state
limit $p_c$ and even beyond\cite{OverJJ2022,Over2023}. 
The central idea is that the pseudogap temperature is the onset of electronic phase separation 
that originates all CDW/CO phenomena and from which we can infer 
a clear intimate relation with the local superconducting amplitude $\Delta_{\rm sc}$, 
like that found by Raman responses\cite{Intimate2019}
of the Hg-1223 compound above $ T_{\rm c}$. Recalling that
charge modulations are ubiquitous present in cuprates; they appear in the
Mott-insulator phase that is precursive to the superconductivity\cite{Kang2023B}, in 
all levels of doped superconductors\cite{Comin2016,Over2023}, and, as recently observed,
in thin LSCO films\cite{OverJJ2022} up to $p \approx 0.35$. This last result 
may appear to be beyond the pseudogap limit but it is in agreement with
the temperature-independent Hall coefficient which is one way to define the
pseudogap temperature $T^*$\cite{Hall.1994}.

Therefore, old, new, and recent data suggest an intimate link between charge inhomogeneity
and local superconducting amplitude\cite{Intimate2019}. Despite the pseudogap and
the charge inhomogeneity being much weaker in the overdoped than in the underdoped 
regime, we show that, according the recendt experiments\cite{OverJJ2022,Over2023}
the transition to the superconducting state occurs by the same mechanism:
The average Josephson coupling between the superconducting 
amplitudes in the CDW/CO regions yields long-range order
and superconducting critical temperature $T_{\rm c}$, unifying the long-range
superconducting transition mechanism of under\cite{Mello2020a,Mello2023} and overdoped cuprates.

We emphasize that the superconducting order parameter is developed
in the charge modulation domains, a scenario which is
verified by the STS\cite{Over2023} and STM/susceptibility experiments\cite{OverJJ2018,OverJJ2022}
and which is, within our model, the most important
and basic property of cuprates. The main advantage of our approach concerning
theoretical predictions for strongly overdoped cuprates\cite{Spivak2008}
mentioned by Ref. \onlinecite{OverJJ2022} is that
the above calculations provide a unified model
for the superconducting state of either under or overdoped compounds.

\section{acknowledgements}

We acknowledge partial support from the Brazilian agencies CNPq and 
by Funda\c{c}\~ao Carlos Chagas Filho de Amparo
Pesquisa do Estado do Rio de Janeiro (FAPERJ), Projects No.
E-26/010.001497/2019 and No. E-26/211.270/2021.

%


\end{document}